\documentclass[
    ,final            
  ]
  {aipproc}

\layoutstyle{6x9}

\newcommand{\imag}{\hbox{Im}\,}
\newcommand{\pepe}{\hbox{P.P.}\,}
\newcommand{\dd}{\hbox{d}\,}
\begin{document}

\title{Precise analysis of $\pi\pi$ scattering data
from Roy equations and forward dispersion relations}

\classification{13.75.Lb,11.55.-m,11.55.Fv,11.80.Et}
\keywords      {Pion scattering, dispersive theory}

\author{J.R. Pel\'aez
\footnote {To appear in the proceedings of the Workshop 
on Scalar Mesons and Related Topics, Lisbon, Portugal, 11-16 Feb 2008.   }
}{
  address={Departamento de F\'{\i}sica Te\'orica~II,
Facultad de Ciencias F\'{\i}sicas,
Universidad Complutense de Madrid,
E-28040, Madrid, Spain.}
}

\author{R. Garc\'ia-Mart\'{\i}n}{
  address={Departamento de F\'{\i}sica Te\'orica~II,
Facultad de Ciencias F\'{\i}sicas,
Universidad Complutense de Madrid,
E-28040, Madrid, Spain.}
}

\author{R.~Kami\'nski}{
   address={Department of Theoretical Physics
Henryk Niewodnicza\'nski Institute of Nuclear Physics,
Polish Academy of Sciences,
31-342, Krak\'ow, Poland.}
 }

\author{F.J. Yndur\'ain}{
   address={Departamento de F\'{\i}sica Te\'orica, C-XI,
  Universidad Aut\'onoma de Madrid,
  Canto Blanco,
 28049, Madrid, Spain.}
 }

\begin{abstract}
We review our recent analysis of $\pi\pi$ scattering data
in terms of Roy equations and Forward Dispersion Relations,
and present some preliminary results in terms of a new set
of once-subtracted coupled equations for partial waves.
The first analysis consists of independent fits to the different 
$\pi\pi$ channels that satisfies rather well the dispersive representation.
In the second analysis we constrain the fit with the dispersion relations.
The latter provides a very precise and model independent description of
data using just analyticity, causality and crossing.
\end{abstract}

\maketitle


\section{Introduction}

A precise knowledge of pion-pion scattering is of interest 
since it provides a test of Chiral Perturbation Theory (ChPT)
as well as useful information about 
quark masses and the chiral condensate \cite{Gasser:1983yg}.
The reaction, at least in the elastic regime, is also remarkably 
 symmetric in terms of isospin and crossing symmetries.
Unfortunately, the existing experimental information from $\pi\pi$ scattering has many conflicting data sets at intermediate energies and no data at all close to the interesting threshold region. For many years this fact has
made it very hard to obtain conclusive results on $\pi\pi$ scattering at low energies or in the sigma region. However, recent \cite{Batley:2007zz}
and precise experiments
on kaon decays, related to $\pi\pi$ scattering at very low energies, have renewed
the interest on this process.

The dispersive integral formalism is 
model independent, just based on analyticity and crossing,
and relates the $\pi\pi$ amplitude 
at a given energy with an integral over the whole energy range, increasing the precision 
and providing information
on the amplitude at energies where data are poor,
or in the complex plane.  In addition, 
it makes the parametrization of the data irrelevant once it is included in the integral
and relates different 
scattering channels among themselves. For all these reasons it is well suited to
study the threshold region or the poles in the complex plane associated to resonances
(see H. Leutwyler and R. Garc\'{\i}a Mart\'{\i}n 
\cite{Ruben} talks on this conference and references therein).

Our recent works make use of two complementary 
dispersive approaches, in brief:

$\bullet$ {\it Forward Dispersion Relations (FDRs):} 
They are calculated at $t=0$ so that the 
unknown large-t behavior of the amplitude is not needed.  
We consider two symmetric and one asymmetric isospin combinations, 
to cover the isospin basis.  The symmetric ones,
$\pi^0\pi^+$ and $\pi^0\pi^0$, have two subtractions and
can be written as
\begin{equation}
\real F-F(4M_{\pi}^2)=
\frac{s(s-4M^2_\pi)}{\pi}\pepe\int_{4M_{\pi}^2}^\infty\dd s'\,
\frac{(2s'-4M^2_\pi)\imag F(s')}{s'(s'-s)(s'-4M_{\pi}^2)(s'+s-4M_{\pi}^2)}
\label{FDR1}
\end{equation}
where $F$ stands for the
$F_{0+}(s,t)$ or $F_{00}(s,t)$ amplitudes. All contributions 
to their integrands are positive, which makes them very precise.
The antisymmetric isospin combination $I_t=1$ does not require subtractions:
\begin{equation} F^{(I_t=1)}(s,0)=\frac{2s-4M^2_\pi}{\pi}\,\pepe\int_{4M^2_\pi}^\infty\dd s'\,
\frac{\imag F^{(I_t=1)}(s',0)}{(s'-s)(s'+s-4M^2_\pi)}. 
\label{FDR2}
\end{equation}
We have implemented all of them up to $\sqrt{s}\simeq1420$~MeV

$\bullet$ {\it Roy Equations (RE) \cite{Roy:1971tc}
}: they are an infinite set of coupled equations
fully equivalent to nonforward 
dispersion relations, plus some $t-s$ crossing symmetry, 
 written in terms of partial waves of definite isospin I and angular momentum $l$.
The complicated left cut contribution is rewritten 
as series of integrals over partial waves in the physical region:
\begin{equation}
\real f^{(I)}_l(s)=C_l^{(I)}a_0^{(0)}+{C'_l}^{(I)}a_0^{(2)}
+\sum_{l',I'}
\pepe\int_{4M^2_\pi}^\infty\dd s' 
K_{l,l';I,I'}(s',s)\imag f^{(I')}_{l'}(s').
\label{Roy}
\end{equation}
where the  $C_l^{(I)}$, ${C'_l}^{(I)}$ constants and $K_{l,l';I,I'}$ kernels
are known.  
In practice, the calculation is truncated at $J<2$ and at some cutoff
energy $s_0$. The
$J\geq2$ waves and the high energy are treated as input.
RE are well suited to study poles of resonances but are limited
to $\sqrt{s}\leq8m_\pi\simeq1120\,$MeV. At present, we have implemented them up to $\sqrt{s}\simeq2m_K$.


The use of RE has gained interest with three aims: to improve the precision
of scattering data, to test ChPT, or to use ChPT
to obtain the subtraction constants at low energies, which can be recast in terms of the scattering lengths $a^{(0)}_{0}$ and $a^{(2)}_{0}$, and obtain precise
predictions on $\pi\pi$ scattering. In particular 
a series of RE analysis in \cite{Bern} using $\pi\pi$ data parametrizations
for the $l>2$ waves and above 800 MeV for the rest,
 as well as some Regge input, was performed 
with and without ChPT constraints. 
The latter provided, $a^{(0)}_{0}=0.220\pm0.005\,m_\pi^{-1}$ and $a^{(2)}_{0}=-0.0444\pm0.0010\,m_\pi^{-1}$, an extremely precise claim,
together with predictions for other scattering lengths and the S and P wave phase shifts up to 800 MeV.
Some of the input, particularly the Regge theory and the D waves, was questionable \cite{critica}, but it certainly seems to have
a very small influence in the threshold region of the scalar waves \cite{Caprini:2003ta}.

In recent years the Krakow-Paris \cite{Kaminski:2002pe} and Paris \cite{DescotesGenon:2001tn} groups have performed other RE analysis. The former resolved the long-standing ambiguity, discarding the so-called "up"' solution, including in their analysis an study
using polarized target data. The latter checked the  calculation in \cite{Bern}
and claimed an small discrepancy in the Olsson sum rule. 

\section{Our analysis}
\vspace*{-.3cm}

The approach followed in the series of works by our group \cite{Kaminski:2006qe} can be summarized as follows: (1)
We first obtain simple fits to each $\pi\pi$ channel independently,
that we call "`Unconstrained Data Fits"' (UDF). In this way 
all waves are uncorrelated and can be easily changed if new, more precise data becomes available. This actually happened, for example,
in one of our most recent works \cite{Yndurain:2007qm} where we have included the newest $K_{l_4}$ data \cite{Batley:2007zz},
which is very precise.  
At different stages of our approach we have also fitted
Regge theory to $\pi\pi$ high energy data, and as our precision was improving, we have improved some 
of the UDF fits with more flexible parametrizations.
(2) Next, we check how well these UDF satisfy the dispersion relations. Surprisingly some of the most widely used parametrizations fail to satisfy the FDRs. We then select the data parametrizations in better agreement with FDRs. (3) Finally, we impose in the fits the dispersion relations. This provides very precise data fits where all waves are correlated -- thus we call them "`Constrained Data Fits"' (CDF)-- which are consistent
with analyticity, unitarity, crossing, etc... Initially we only considered FDRs but in our most recent
work, whose results we review next, we have included RE.

In Fig.1 we show our unconstrained fits to data (UFD) for several
waves. The details of their simple parametrizations can be found in \cite{Kaminski:2006qe}. We only plot explicitly the constrained fits (CDF)
for the isospin-2 waves since in all other cases they are indistinguishable from the UDF, showing their remarkable stability. 
Actually, all CDF waves differ from the UDF waves by less than one $\sigma$, except for $l=2$, $I=2$ that deviates by 1.5 standard deviations
from the unconstrained case. In the bottom row of Fig.1 we show in greater detail both the UFD and CFD results for the S0 wave, which is probably very controversial and the relevant one for this conference.

\begin{figure}
\centering
\includegraphics[height=.9\textwidth]{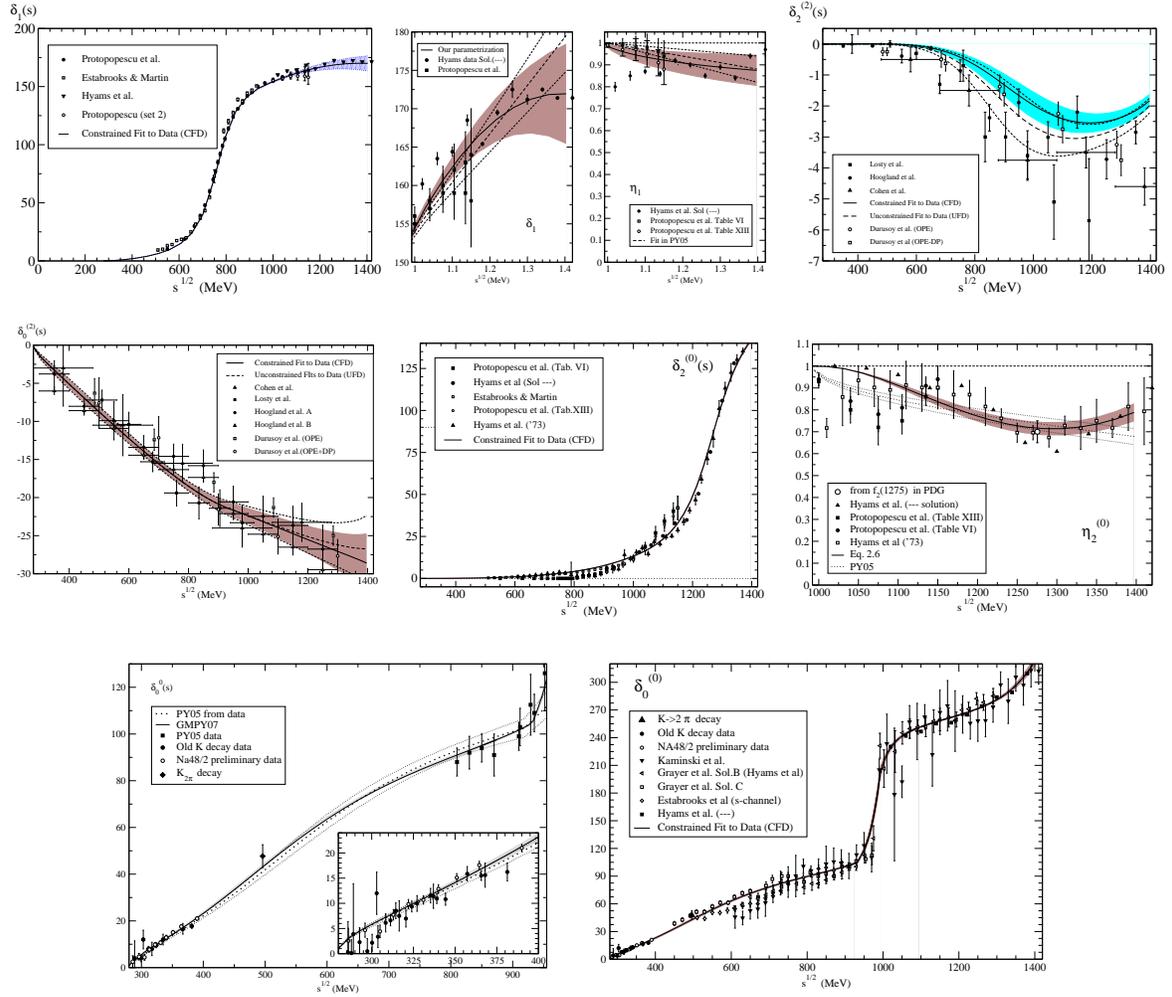}
  \caption{Unconstrained fits to different $\pi\pi$ partial waves.
Almost all constrained fits are indistinguishable except those for 
S2, D2 and S0 waves that are also shown here.}
\end{figure}

In order to quantify how well the dispersion relations are satisfied, we define six $\Delta_i$ as the difference between the left and right sides of each dispersion relation
in Eqs.\eqref{FDR1},\eqref{FDR2} and \eqref{Roy}, whose uncertainties we call $\delta\Delta_i$. Next, we define the average discrepancies
\begin{equation}
\bar{d}_i^2\equiv\frac{1}{\hbox{number of points}}
\sum_n\left(\frac{\Delta_i(s_n)}{\delta\Delta_i(s_n)}\right)^2,
\end{equation}
where the values of $s_n$ are taken at intervals of 25 MeV. 
Note the similarity with an averaged $\chi^2/(d.o.f)$  and thus $\bar{d}_i^2\leq1$
implies fulfillment of the corresponding dispersion relation.
In Table 1 we show the average discrepancies of the UDF 
for each FDR, up to two different energies, and each RE up to $\sim2m_K$. 
Although the total average discrepancy of the UDF set is
practically one, they can be clearly improved in the high energy region
of the antisymmetric FDR and in the scalar isospin-2 RE. This is actually done 
in the CDF set, which is obtained by minimizing:
\begin{equation}
\chi^2\equiv
\left\{\bar{d}^2_{00}+\bar{d}^2_{0+}+\bar{d}^2_{I_t=1}+\bar{d}^2_{S0}+\bar{d}^2_{S2}+\bar{d}^2_P\right\}
W
+\bar{d}^2_I+\bar{d}^2_J+\sum_i\left(\frac{p_i-p_i^{\rm exp}}{\delta p_i}\right)^2.
\end{equation}
where $p_i^{exp}$ are all the parameters of the different UDF parametrization for each wave or Regge trajectory, 
thus ensuring the data description, and
$d_I$ and $d_J$ are the discrepancies for a couple of crossing sum rules. The weight $W=9$ was estimated from
the typical number of degrees of freedom needed to describe the shape of the dispersion relations.

\begin{table}
\begin{tabular}{l|cc||cc}
&\multicolumn{2}{c||}{ Unconstrained Data Fits (UDF)DF}&\multicolumn{2}{c}{Constrained Data Fits (CDF)}\\
\hline
&$s^{1/2}\leq 932\,$MeV&$s^{1/2}\leq 1420\,$MeV&$s^{1/2}\leq 992\,$MeV&$s^{1/2}\leq 1420\,$MeV\\
$\pi^0\pi^0$ FDR& 0.12 & 0.29 & 0.13 & 0.31\\
$\pi^+\pi^0$ FDR& 0.84 & 0.86 & 0.83 & 0.85\\
$I_{t=1}$ FDR& 0.66 & 1.87 & 0.13 & 0.70\\
\hline
&\multicolumn{2}{c||}{$s^{1/2}\leq 992\,$MeV}&\multicolumn{2}{c}{$s^{1/2}\leq 992\,$MeV}\\
Roy eq. S0&\multicolumn{2}{c||}{0.54}&\multicolumn{2}{c}{0.23}\\
Roy eq. S2&\multicolumn{2}{c||}{1.63}&\multicolumn{2}{c}{0.25}\\
Roy eq. P &\multicolumn{2}{c||}{0.74}&\multicolumn{2}{c}{0.002}\\
\hline 
\end{tabular}
\caption{Average discrepancies $\bar d^2$ of the UDF and CDF for each 
FDR and RE. On average, the UDF are 
consistent with dispersion relations. Note the remarkable CDF consistency.
\vspace*{.1cm}}
\end{table}

From the Table it is clear that the CDF set satisfies remarkably 
well all dispersion relations within uncertainties, and 
hence can be used directly and inside the Olsson sum rule to obtain the following precise determination
{\it from data}: $a^{(0)}_{0}=0.223\pm0.009\,m_\pi^{-1}$ and $a^{(2)}_{0}=-0.0444\pm0.0045\,m_\pi^{-1}$. This is in remarkable agreement with the predictions of RE and ChPT of \cite{Bern}.
Nevertheless, the agreement is fairly good only up to roughly 450 MeV,
but from that energy up to 800 MeV those predictions 
deviate from our data analysis. We should stress that we are nevertheless talking about a disagreement of a few degrees at most
and would affect the determination of the sigma mass by tens of MeV at most, which is a remarkable improvement compared with the situation just a few years ago and the
huge and extremely conservative uncertainties quoted in the PDG for the $\sigma$ mass and width, of hundreds of MeV.

The other waves are of less relevance for this conference and we comment them very briefly, since the details can be found in \cite{Kaminski:2006qe}. The best determination of threshold parameters is obtained by using the CDF set inside appropriate sum rules \cite{Kaminski:2006qe}. 
In brief, we agree with \cite{Bern} in the P-wave scattering length, but find disagreements of 2 to 3 standard deviations in the P-wave slope, and also in some $D$ wave parameters. 

In summary the CDF set provides a model independent and very precise description
of the $\pi\pi$ scattering data consistent with analyticity and crossing. 

\begin{figure}
\centering
\includegraphics[height=.9\textwidth]{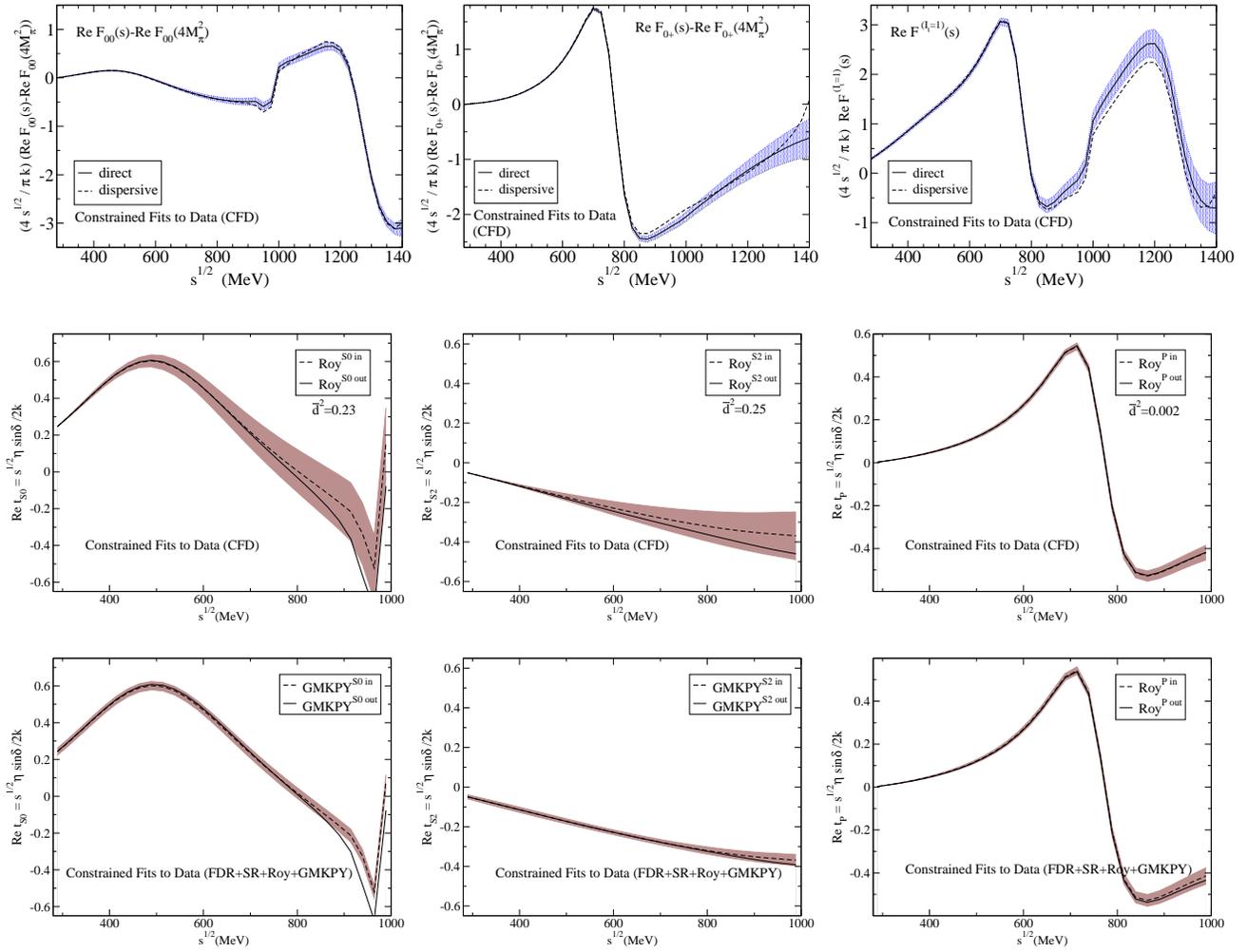}
  \caption{Fulfillment of Forward Dispersion Relations (upper row) and RE (middle row) for the CDF set. In the lower row we show preliminary fits constrained
  to our modified set of once-subtracted Roy-like equations (GMKPY). Although we expect to improve the central values, these plots illustrate the remarkable improvement in the
  resulting uncertainties.}
\end{figure}


\vspace{-4mm}
\section{Outlook}

\vspace{-3mm}
The upper row plots in Fig.2 show that, despite the small uncertainties, the CDF set satisfies remarkably well the FDRs
up to 1420 MeV.  The same happens for the RE, although now the uncertainties,
small close to threshold, become, for the scalar channels, rather large around 800 MeV.
This is due to the polynomial factors in the RE that multiply the scalar scattering lengths, particularly $a^{(2)}_0$.  In order to avoid this, we have derived a modified set of 
once-subtracted Roy-like equations (GMKPY for brevity). 
Actually, the plots of the bottom row of Fig.3 show our preliminary results for a constrained
fit including also the GMPKY. Note that close to threshold the resulting GMKPY uncertainties are larger than for the standard RE, but the uncertainties are dramatically reduced at intermediate energies, the region of interest for scalar spectroscopy and this conference (see the talk by R. Garc\'{\i}a-Mart\'{\i}n). Obviously, these tiny uncertainties will force us to improve
the region around 850 to 1 GeV with more flexible parametrizations, and in particular the matching of the low and intermediate energy regions for the scalar-isoscalar channel,  not sufficiently smooth at the moment (see Fig 1, bottom). Work is in progress in these directions.

\vspace{-5mm}
\begin{theacknowledgments}
J.R.P. thanks the organizers for creating such a stimulating conference. 
\end{theacknowledgments}

\end{document}